\documentstyle[aps,epsf,psfig,epsfig,xspace,]{revtex}  % DO NOT DELETE 

\begin{document}        %  DO NOT DELETE OR CHANGE THIS LINE
\setlength{\voffset}{1.7cm}

\title{
\begin{flushleft}
\center {\normalsize{DPF 2003: Annual Meeting of the Division of Particles and Fields of
the American Physical Society,\\ 5-8 April 2003, Philadelphia, Pennsylvania}}
\end{flushleft}
\begin{flushleft}
\end{flushleft}
\bf \boldmath Study of the Absolute Branching Fraction of
{\boldmath{$\Upsilon(4S) \rightarrow B^0 \bar {B}^{0}$}} 
with Partial Reconstruction of {\boldmath{$\bar {B}^0 \rightarrow D^{*+} 
\ell^{-} \bar{\nu}_{\ell}$}} at \boldmath{\bf BABAR}
}
\author{Romulus Godang and Donald Summers}
\address{University of Mississippi-Oxford}

\maketitle              % Creates the title area, Do Not Remove

\vspace*{0.3cm}

\begin{abstract}        % Do Not Delete this line
Based on a data sample of 81.7 fb$^{-1}$ collected at the $\Upsilon(4S)$
resonance with the {\sc BaBar} detector at the PEP-II asymmetric-energy $B$ 
Factory at SLAC, 
we present the current status of the measurement of the branching fraction of 
$\Upsilon(4S) \rightarrow B^0 \bar {B}^{0}$.   Our study of the decay was
performed through the exclusive decays $\bar {B}^0 \rightarrow D^{*+} \ell^{-} 
\bar{\nu}_{\ell}$ using a partial reconstruction method, where the $D^{*+}$ is 
detected only through the soft pion daughter from the decay 
$D^{*+} \rightarrow D^{0} \pi^{+}$.
\end{abstract}   	% Do Not Delete this line

\section{Introduction}               
The exclusive decay $\bar{B}^0 \rightarrow D^{*+} \ell^- \bar{\nu}_{\ell}~ 
(D^{*+}\rightarrow D^0 \pi^+)$ has been previously 
analyzed using a partial reconstruction technique~\cite{godang02}.  
In this technique,
the $D^{*+}$ is identified without reconstructing the $D^{0}$ 
meson, and the presence of an undetected neutrino is inferred by conservation 
of momentum and energy.  This approach is possible due to the extremely
low decay energy of this mode.  The soft $\pi^{+}$ carries sufficient
information to determine an approximate four-momentum of the $D^{*+}$ meson.
The partial reconstruction technique may result in a gain
of as much as a factor of 10 in statistics compared
to the full reconstruction technique, though the effective gain may be
less due to background.

This study can improve the understanding of the branching fraction of all 
$B$ decays measurements, including studies of CP violation and 
the Cabibbo-Kobayashi-Maskawa quark-mixing matrix element,
$V_{cb}$.  Furthermore, this study can be a significant contribution to 
enhance our knowledge of isospin violation in $\Upsilon(4S)$ decay.
The isospin violation is due to the mass difference of the $u$ and $d$ quarks 
and due to electromagnetic interactions.  
All currently published branching fraction measurements based on an admixture 
of $B$ mesons at the $\Upsilon(4S)$ assume that ${\cal B}(\Upsilon(4S) \rightarrow B \bar B) 
\approx 100\%$~\cite{pdg2002}.
All measurements of fundamental parameters at the $\Upsilon(4S)$ 
are limited by the uncertainty on the ratio of the branching fraction of 
$\Upsilon(4S) \rightarrow B^+ \bar {B}^{-}$ to the branching 
fraction of $\Upsilon(4S) \rightarrow B^0 \bar {B}^{0}$ defined as
${f_{+-} \over f_{00}} \equiv R^{+/0}$.
The current measurements of $f_{+-}\over f_{00}$ are limited its ratio 
within an uncertainty of $8\%$~\cite{silvia}.  
Its values depend on the ratio of the charged and neutral $B$ meson 
lifetime as well as the assumption of 
isospin symmetry.

The $B$ meson velocity in the $\Upsilon(4S)$ rest frame is 
relatively small.  Thus $\beta=$V$/c$:
\begin{eqnarray}
\beta=\sqrt{1-\frac{4m_{B}^{\,2}}{m_{\Upsilon(4S)}^{\,2}}}\approx 0.065, 
\label{eq:b_velocty}
\end{eqnarray}
\noindent
where $m_{B}$ and $m_{\Upsilon(4S)}$ are the masses of the $B$ meson and 
the $\Upsilon(4S)$ resonance, respectively.
Since the final state $B$ mesons are non-relativistic
and have low momentum, it was suggested that the final state interactions
of the $B$ meson can be treated using non-relativistic field theory combined 
with chiral perturbation theory.  One can also calculate the dominant Coulomb 
correction using non-relativistic time-dependent perturbation theory.
The current theoretical predictions of $f_{+-} \over f_{00}$ has 
a variation effect and its ratio range from 1.03 to 1.25~\cite{eichten}:
\begin{eqnarray} 
{f_{+-} \over f_{00}} \equiv {R^{+/0}} \equiv 
\frac{\Gamma(\Upsilon(4S)\rightarrow B^+ B^-)}
{\Gamma(\Upsilon(4S)\rightarrow B^{0}\bar B^{0})} 
\approx 1.03-1.25.
\label{eq:eichten}
\end{eqnarray}

\begin{figure}[!htb]
\begin{center}
\includegraphics[height=4.5cm]{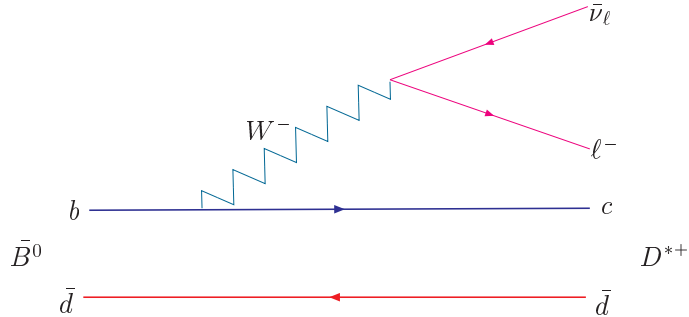}
\caption{Quark-level Feynman diagram for spectator $B$ decays.}
\label{fig:fm_plot}
\end{center}
\end{figure}

The exclusive decay of $B \rightarrow D^{*} \ell \nu$ has the largest 
branching fraction of any exclusive $B$ decay and a relatively
simple theoretical interpretation. 
Figure~\ref{fig:fm_plot} shows the spectator quark-level Feynman diagram, where the heavy 
$b$ quark decays to either a $c$ or $u$ quark and a lepton pair 
created from the virtual $W$ boson.  The flavor of the $B$ meson 
can be indicated from the charge of the lepton.  
More precisely, a positively charged lepton indicates a $B$ meson 
with a $\bar b$ quark, whereas a negatively charged lepton 
indicates a $\bar B$ meson with a $b$ quark. 

\begin{figure}[!htb]
\begin{center}
\includegraphics[height=3.9cm]{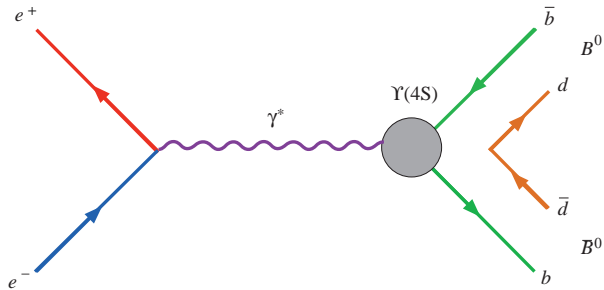}
\caption{The mechanism for $B^0 \bar {B}^{0}$ production in 
        $e^{+} e^{-}$ collisions.}  
\label{fig:upsilon}
\end{center}
\end{figure}

The main focus of this paper is to describe the procedure used in determining 
the absolute branching fraction of $\Upsilon(4S) \rightarrow B^0 \bar 
{B}^{0}$, $f_{00}$, with partial reconstruction of $\bar {B}^0 
\rightarrow D^{*+} \ell^{-} \bar{\nu}_{\ell}$.  
The mechanism for $B^0 \bar {B}^{0}$ production in $e^{+}e^{-}$ collisions 
at the $\Upsilon(4S)$ resonance is shown in Fig.~\ref{fig:upsilon}. 

\section{Data Sample and Event Selection}

The data used in this paper were collected with the {\sc BaBar} detector at 
the PEP-II  asymmetric-energy $B$ Factory at SLAC.  
The ARGUS~\cite{argus87} and UA1~\cite{ua187} discovery of $B^0$ mixing, made possible by the
surprisingly massive top quark~\cite{cdf95}, provided $B^0$ decays that could interfere.
Plans were soon begun to construct an asymmetric B Factory based on PEP to
search for CP violation~\cite{oddone}.  This has now led to a data sample 
of 81.7 fb$^{-1}$ at the $\Upsilon(4S)$ resonance (on-resonance) and 
9.6 fb$^{-1}$ below the resonance (off-resonance).  The total number
of the simulated generic Monte Carlo (MC) for $B^0 \bar {B}^{0}$ and
$B^+ \bar {B}^{-}$ events are equivalent to about 160 fb$^{-1}$ for each pair.

Hadronic events are selected by requiring at least four tracks from 
charged hadrons reconstructed by the Silicon Vertex Detector (SVT)~\cite{burchat94}
and the Drift Chamber (DCH)~\cite{burchat92}.  
This selection helps to remove Bhabha and muon pair events.  
All lepton candidates (electrons and muons) are required to have momenta 
between 1.5~GeV/$c$ and 2.3~GeV/$c$ to suppress the leptons
from the other charm decays.  Soft pion candidates were selected among all
the charged particles with momenta between 60~MeV/$c$ and 200~MeV/$c$.

Electrons are identified by exploiting information from the Electromagnetic 
Calorimeter (EMC)~\cite{barlow99}. 
The ratio $E/p$ is used to provide a good discrimination between electrons and other charged
particles species. $E$ is the measured energy of a shower in the calorimeter 
and $p$ is the measured momentum of the corresponding charged track.  
The efficiency for electrons in the acceptance of the electromagnetic 
calorimeter is $90\%$, with a hadron mis-identification probability of 
less than $1\%$.
We also require the measurements of the specific ionization ($dE/dx$) from 
the drift chamber as well as an information from the Detection of Internally
Reflected Cherenkov ring imaging detector (DIRC)~\cite{schwiening03}.

Muons are identified in the Instrumented Flux Return (IFR)~\cite{anulli02}.  
We use a ``Tight Muon'' selection which provided an efficiency of
about $70\%$ with a hadron mis-identification 
probability about $2\%$.   
Kaons are rejected using information from the Cherenkov light emission in the
DIRC by requiring the consistency with the kaon hypothesis to be
smaller than $5\%$.  More details of the {\sc BaBar} detector are described elsewhere~\cite{nim}.

\section{Analysis Method}

To partially reconstruct $\bar B^{0} \rightarrow D^{*+} \ell^{-} 
\bar\nu_{\ell}$, lepton candidates are combined with soft charged pions from
the decay $D^{*+} \rightarrow D^0 \pi^+$.  The $D^*$ is just massive enough 
to create a $D$ meson and a soft pion.  These two daughters therefore have 
very little momentum in the $D^*$ rest frame.
This pion is often referred to as the ``soft pion,'' and its direction 
coincides approximately with the direction of the parent $D^*$.  
This condition allows us to do an approximation of the $D^*$ four-vector
by measuring only the pion four-vector momentum, without the $D$ meson
reconstruction.

The approximate four-momentum of the $D^*$, 
($\widetilde{E}_{{D^*}},\widetilde{\bf{p}}_{{D^*}}$),
is calculated by scaling the soft pion momentum:
\begin{eqnarray}
E_{D^*} &\simeq& {E_{\pi}\over E^{CM}_{\pi}}m_{D^*} \equiv \widetilde{E}_{D^*},\ {\rm and}\\
{\bf{p}}_{D^*}&\simeq &
{\hat{\bf{p}}}_\pi\times{\sqrt{\widetilde{E}_{D^*}^2-m_{D^*}^2}}
\equiv\widetilde{\bf{p}}_{D^*},
\end{eqnarray}
where $E_\pi$ is the pion energy, $E^{CM}_{\pi} \approx 145$~MeV 
is the energy of the pion in the 
$D^{*}$ center of mass frame, and $m_{D^*}=2.01$~GeV/$c^2$ is the mass of 
the $D^{*}$.

Since the $B$ meson has a very small momentum, one may use an 
approximation for the missing mass squared
($\widetilde {{\cal M}_{\nu}}^{\,2}$) with $|\vec{P}_B| = 0$: 
\begin{eqnarray}
\widetilde{\cal M}_\nu^2 \equiv (E_{\mbox{beam}}-\widetilde{E}_{{D^*}} - 
E_{\ell})^2-(\widetilde{\bf{p}}_{{D^*}} + {\bf{p}}_{\ell})^2\ .
\label{eqn:mms}
\end{eqnarray}
The $\widetilde {{\cal M}_{\nu}}^{\,2}$ distribution will peak near zero 
if the decay has been properly reconstructed and the neutrino is 
the only missing particle.  

\section{Tag Events Selection}

The 'single tag' events are referred at least one neutral $B$ partially 
reconstructed through the exclusive decays $\bar {B}^0 \rightarrow D^{*+} 
\ell^{-} \bar{\nu}_{\ell} (D^{*+}\rightarrow D^0 \pi^+)$.  The total signal 
yield of these events is denoted as $N_{s}$.  The 'double tag' events are 
referred to two neutral $B$ partially reconstructed in the same 
decay channel as above.  These events are obtained by reconstructing 
the other $B$ mesons within the single tag events.  Its total signal yield 
is denoted as $N_{d}$.

The total numbers of the single tag and the double tag reconstructed 
signal events, $N_{s}$ and $N_{d}$, have the following relationships 
to the branching fractions respectively:
\begin{eqnarray}
N_{s} & = &  
      2 \times N_{B\bar{B}} f_{00} \times 
      {\cal B}(\bar{B}^0 \rightarrow D^{*+} \ell^- \bar{\nu}_{\ell}) 
      \times {\cal B}(D^{*+} \rightarrow D^0 \pi^+) \times 
      \epsilon_{0+}
\label{eq:nsingle}
\end{eqnarray} 
\begin{eqnarray}
N_{d} & = &  
      N_{B\bar{B}} f_{00} \times 
      [{\cal B}(\bar{B}^0 \rightarrow D^{*+} \ell^- \bar{\nu}_{\ell}) 
      \times {\cal B}(D^{*+} \rightarrow D^0 \pi^+) \times 
      \epsilon_{0+}]^{2}
\label{eq:ndouble}
\end{eqnarray} 
where $f_{00}$ is the fraction of neutral $B$ mesons in $\Upsilon(4S)$ events,
and $\epsilon_{0+}$ is the reconstruction efficiency for the respective 
$B\rightarrow D^*\ell\nu~(D^{*+}\rightarrow D^{0}\pi^{+})$ modes. 

Combining these yields, one solves for $f_{00}$
\begin{eqnarray}
f_{00} & = &  
       N_{s}^{2} \over {4 N_{d} N_{B \bar{B}}} 
\label{eq:f00}
\end{eqnarray}
where $N_{B \bar{B}}$ is the total numbers of $B \bar{B}$ events.
This study has several advantages both over the measurements with fully 
reconstructed $B$ mesons and the measurements of the ratio of charged 
to neutral production of $B$ mesons at the $\Upsilon(4S)$ resonance.  
First, a lower systematic error may be obtained since this method is 
independent of the branching fraction on the $D^{*}$ decays with
its large uncertainty.  
Secondly, by combining it with another direct measurement of $f_{+-}$ 
this method could address the question of additional substantial 
$\Upsilon(4S)$ decay modes~\cite{physbook}.

\section{Backgrounds}

The continuum background events are non-resonant decays of $e^{+} e^{-} 
\rightarrow \gamma^* \rightarrow q\bar q$ where $q = u, d, s, c$.
These backgrounds are generally collimated into two back-to-back jets 
events while \mbox{$\Upsilon(4S) \rightarrow B\bar B$} events 
are much more isotropic in the $\Upsilon(4S)$ rest frame.
To reduce the continuum background events, the ratio $R_2=H_2/H_0$ of 
Fox-Wolfram moments has been used~\cite{wolfram}.  This variable can take 
values between 0 and 1, and tends toward higher values for jet-like events 
and lower values for events with isotropic distributions of final state 
particles.  By requiring $R_2 < 0.4$, we retain nearly all $B\bar B$ events 
while reducing the contribution from $q\bar q$ continuum by approximately 50\%.

The contribution from the continuum background events are estimated 
by analyzing the off-resonance data and scaling the result to correct for 
differences in the integrated luminosity and center of mass energy.
The average continuum scaling factor, $\lambda_{\it continuum}$, is
\begin{eqnarray}
\lambda_{\it continuum} = \frac{\cal L_{\it on}}{\cal L_{\it off}} 
\frac{E_{\it off}^{\,2}}{E_{\it on}^{\,2}} = 8.44. 
\end{eqnarray}

\begin{figure}[!htb]
\vspace*{-3.0cm}
\begin{center}
\includegraphics[height=16cm,width=12cm]{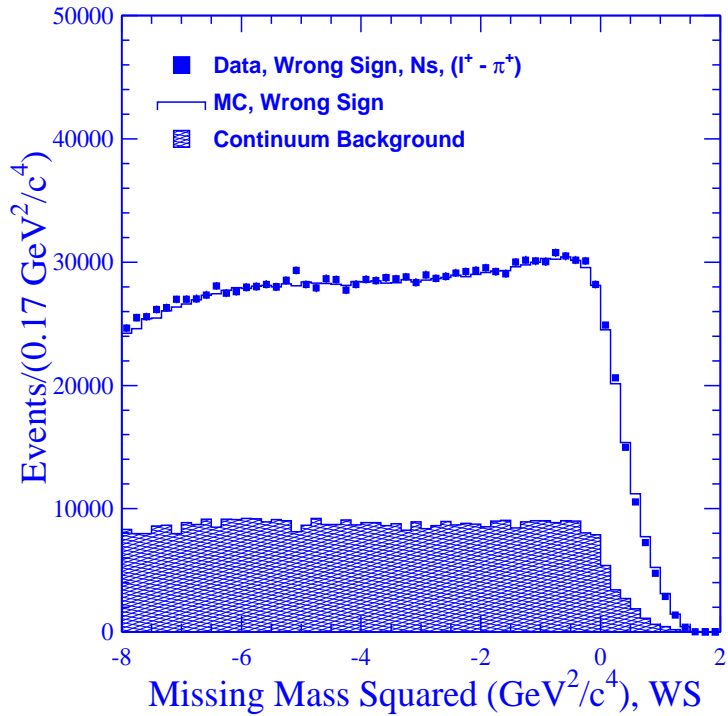}
\vspace*{-2.0cm}
\caption{The single tag distribution in $\widetilde{\cal M}_\nu^2$ for 
         the wrong-sign $(\ell^{+}-\pi^+)$ candidates. The plot shows 
         data on resonance (dotted), Monte Carlo normalized to data 
         in the sideband region (solid histogram) and the scaled continuum 
         background (hatched).  
         Note: the scaled continuum background has been subtracted from the 
         data.}
\label{fig:wrongsign}
\end{center}
\end{figure}

The combinatoric background, also known as uncorrelated background, 
for single tag events is defined as a random combination of real leptons 
from $B$ decays that are paired with right sign soft pions that come from 
the other $B$.  This background can also be due to the low momentum soft 
pions not necessarily coming from a $D^{*}$, produced by either the same or 
other $B$.  The combinatoric background events are estimated using 
Monte Carlo data simulation (MC).
Its normalization is obtained by fitting the Monte Carlo 
simulation to the data in the sideband region, 
$-8<\widetilde{\cal M}_\nu^{\,2}<-4~$GeV$^{2}/c^4$.
The overall $\widetilde{\cal M}_\nu^2$ 
distribution of the combinatoric background is dominated by phase space, 
i.e., a hard lepton and soft pion distributed isotropically will produce 
a distribution similar to signal events.  A simple overall test of 
its reliability is the counting of the wrong sign ($\ell^{+}-\pi^{+}$) 
candidates, where no signal is expected as shown in Fig.~\ref{fig:wrongsign}.

\begin{figure}[!htb]
\vspace*{-4.0cm}
\begin{center}
\includegraphics[height=22cm,width=18cm]{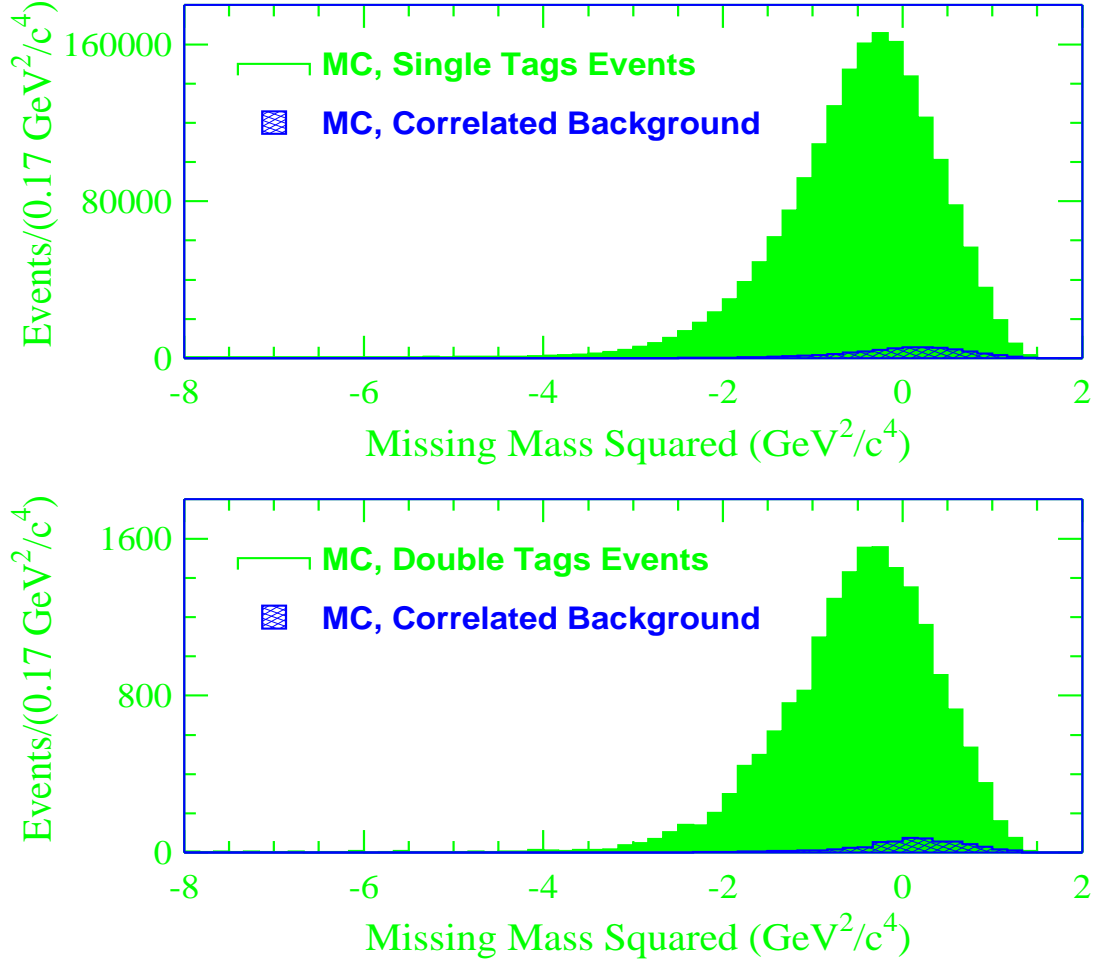}
\vspace*{-3.0cm}
\caption{The right sign data distributions in $\widetilde{\cal M}_\nu^2$
         of Monte Carlo data simulation.  
         The upper plot shows the single tag events (solid histogram) 
         and correlated background (shaded). 
         The lower plot shows the double tag events (solid histogram) 
         and correlated background (shaded).}
\label{fig:dss_signal}
\end{center}
\end{figure}

The correlated background is obtained from the right sign combination of
leptons with soft pions either from the same or different $B$, however, 
the soft pions come from $B \rightarrow D^{*}\pi \ell \bar{\nu}_{\ell}$.
The relatively high momentum cut of
1.5~GeV/$c$ on the leptons is chosen to reduce the contribution of 
the correlated background, which is believed to be small but it is
otherwise difficult to separate them kinematically from the tag
signal events in $\widetilde{\cal M}_\nu^{\,2}$ distribution.
The single tag and double tag events accumulate in the signal region, 
$\widetilde {{\cal M}_{\nu}}^{\,2} > -2.0~$GeV$^{2}/c^4$, as shown in 
Fig.~\ref{fig:dss_signal}.
All contributions of the correlated background come from the decays of 
the type $\bar{B} \rightarrow D^{*} \pi \ell^- \bar{\nu}_{\ell}$ through 
the channels:
\begin{eqnarray*}
B^- &\rightarrow &D^{*+} \pi^- \ell^- \bar{\nu}_{\ell}\\
\bar{B}^0 &\rightarrow& D^{*0} \pi^+ \ell^- \bar{\nu}_{\ell} \\
B^- &\rightarrow & D^{*0} \pi^0 \ell^- \bar{\nu}_{\ell} \\
\bar{B}^0 &\rightarrow &D^{*+} \pi^0 \ell^- \bar{\nu}_{\ell}\ ,
\end{eqnarray*}
where $D^*\pi$ may or may not be from an excited charm resonance 
such as the $D_{1}(2420)^{0}$~\cite{argus86} and the additional pion is 
not detected. However, only the modes with $D^{*+}$  could contribute to 
$\widetilde{\cal M}_\nu^2$.  
These modes has been extensively studied by the ARGUS, CLEO and {\sc BaBar} 
Collaborations~\cite{godang02}.

\section{Signal Yields}

\begin{figure}[!htb]
\vspace*{-4.0cm}
\begin{center}
\includegraphics[height=22cm,width=18cm]{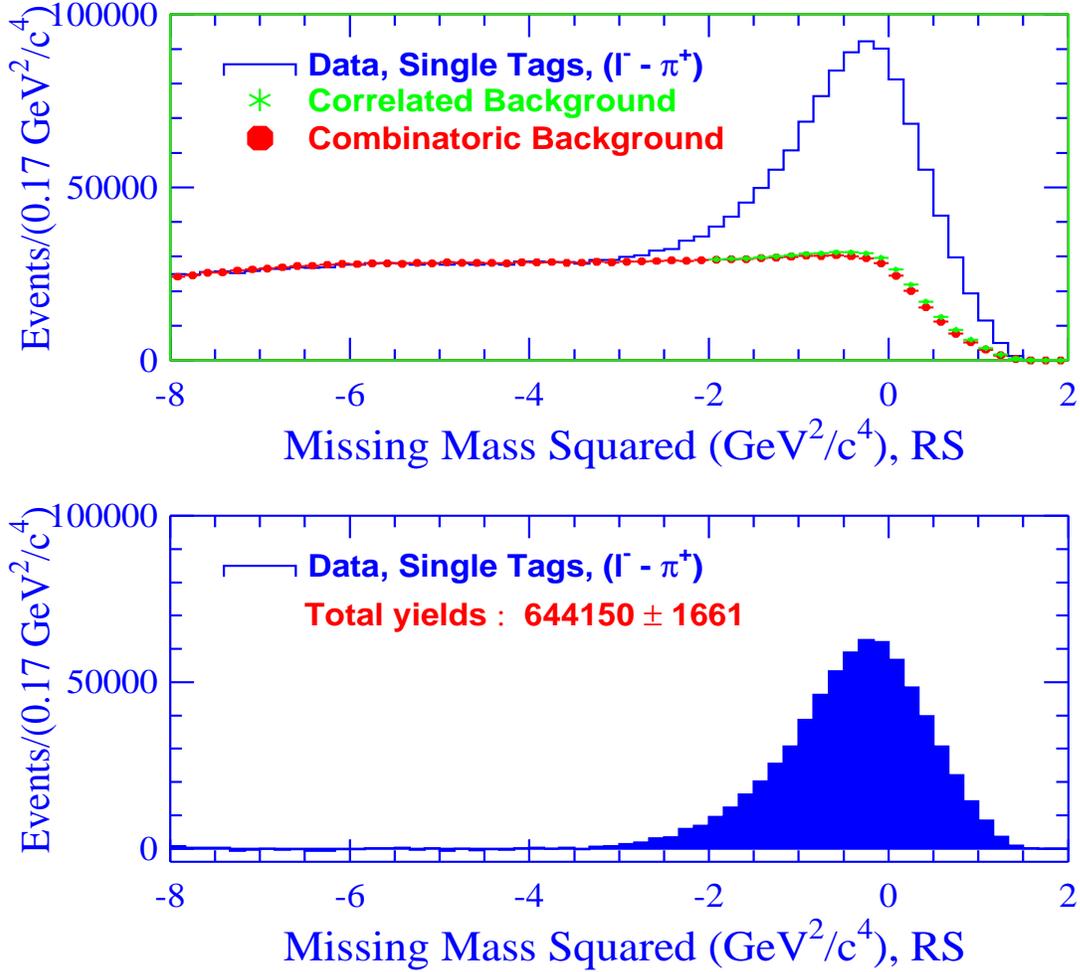}
\vspace*{-3.0cm}
\caption{The single tag yield in $\widetilde{\cal M}_\nu^2$. 
         The upper plot shows the right sign data with the continuum background
         subtracted, correlated background (asterisk) and the combinatoric 
         background (dotted) estimated by Monte Carlo simulation. 
         The lower plot shows the total signal yield of the single tag 
         candidates after all backgrounds are subtracted.}
\label{fig:single_yields}
\end{center}
\end{figure}

\begin{figure}[!htb]
\begin{center}
\vspace*{-3.0cm}
\includegraphics[height=22cm,width=18cm]{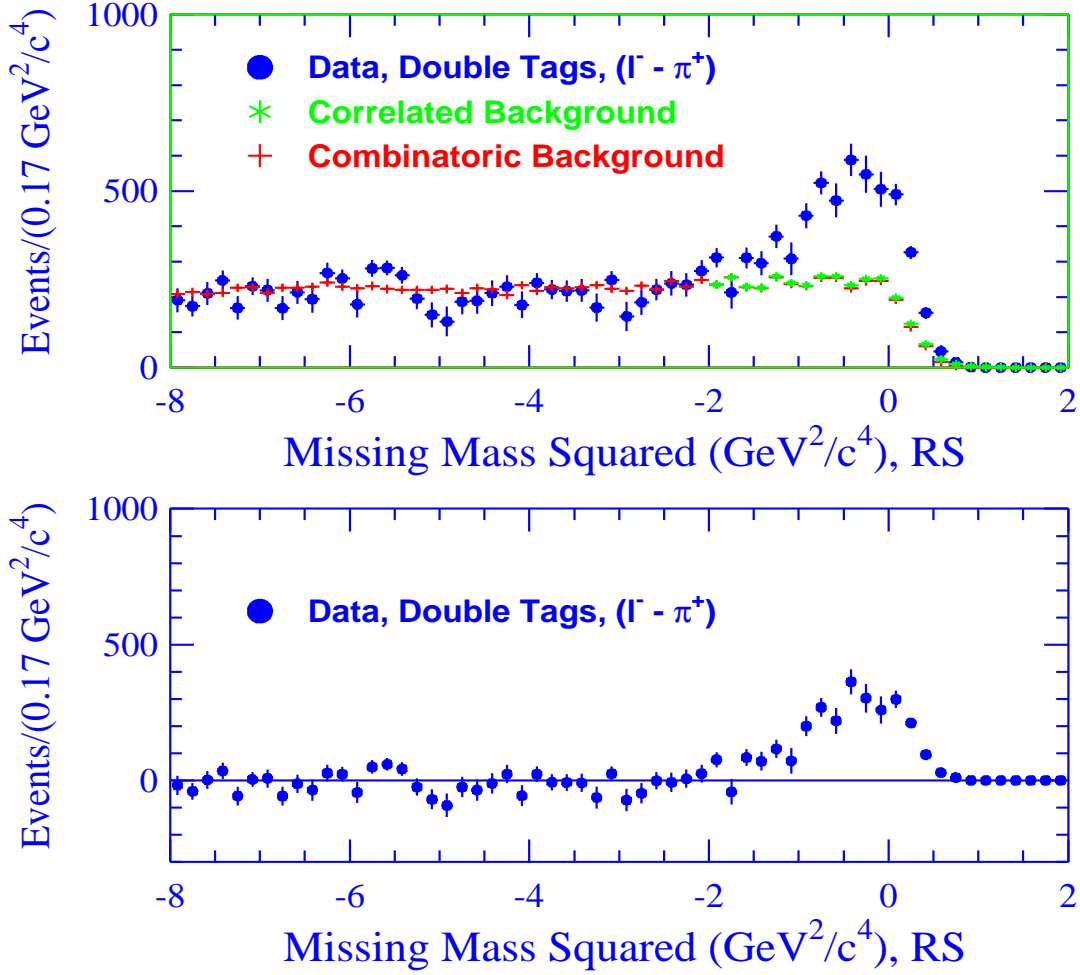}
\vspace*{-3.0cm}
\caption{The double tag yield in $\widetilde{\cal M}_\nu^{\,2}$. 
         The upper plot shows the right sign data, correlated background 
         (asterisk) and the combinatoric background (plus sign) estimated 
         by Monte Carlo simulation.
         The lower plot shows the total signal yield of the double tag 
         candidates after all backgrounds are subtracted.}
\label{fig:double_yields}
\end{center}
\end{figure}

After all background subtractions described above, the total signal yield 
for the single tag and double tag events are extracted by counting 
($\ell^{-}-\pi^{+}$) candidates which fall in the signal region, 
$\widetilde{\cal M}_\nu^{\,2}>-2~$GeV$^{2}/c^{4}$.  
Figure~\ref{fig:single_yields} and  Figure~\ref{fig:double_yields} 
show the total signal yield for the single tag and double tag events
respectively.

\begin{figure}[!htb]
\vspace*{-10.0cm}
\begin{center}
\includegraphics[height=24cm,width=18cm]{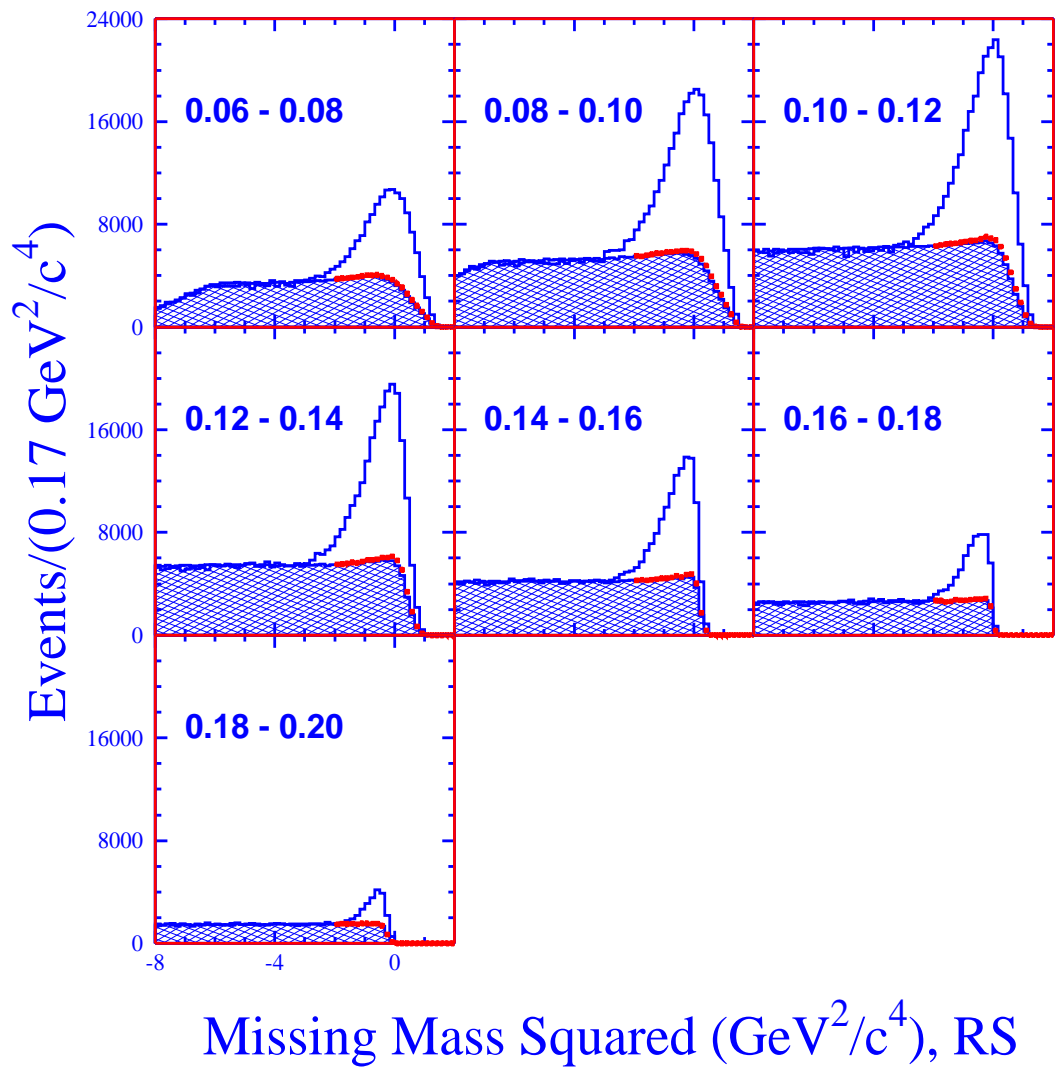}
\vspace*{-3.3cm}
\caption{The single tag distributions in $\widetilde{\cal M}_\nu^2$ 
         for 20~MeV/$c$ bins of soft pion momentum.  
         The plot shows the right sign data (solid histogram), 
         the combinatoric background (shaded) and correlated 
         background (dotted) estimated by Monte Carlo simulation.}
\label{fig:single_mombin}
\end{center}
\end{figure}

In order to have a better understanding of the Monte Carlo simulation 
modeling, the signal yield of the single tag events is extracted in bins 
of 20~MeV/$c$ in the soft pion momentum.  There is an excellent agreement 
between the data of the combinatoric background and Monte Carlo simulation 
in the sideband region, $-8<\widetilde{\cal M}_\nu^{\,2}<-4~
$GeV$/c^{4}$, as shown in Fig.~\ref{fig:single_mombin}.

\section{Conclusion}

This analysis will be the first measurement of the absolute branching 
fraction of $\Upsilon(4S) \rightarrow B^0 \bar {B}^{0}$
with partial reconstruction of $\bar {B}^0 \rightarrow D^{*+}
\ell^{-} \bar{\nu}_{\ell}$.  It is a direct experimental measurement of 
$\Upsilon(4S) \rightarrow B^0 \bar {B}^{0}$ that is independent of 
$\bar {B}^{0}$ lifetime as well as the branching fractions of 
$\bar {B}^{0}$ and $D^{*+}$.
The currently published measurements of 
the ratio of charged to neutral production of $B$ mesons at 
the $\Upsilon(4S)$ resonance, $f_{+-}\over f_{00}$, is consistent with 
unity within an error of $8\%$~\cite{silvia}.

By comparing the number of events with both one and two reconstructed
$\bar {B}^0 \rightarrow D^{*+} \ell^{-} \bar{\nu}_{\ell}$ candidates,
the absolute branching fraction of $\Upsilon(4S) 
\rightarrow B^0 \bar {B}^{0}$ will be obtained.
The analysis is currently under review and will be published soon.  
The expected statistical uncertainty is less than $6\%$.
The major systematic uncertainties for the $f_{00}$ value have been studied,
for example, the systematic error due to the $B$ counting is about
$1.1\%$~\cite{chris}.  It is mainly due to the uncertainties in the tracking 
efficiency.

\section{Acknowledgments}

We are grateful for the extraordinary contributions of our PEP-II 
colleagues in achieving the excellent luminosity and machine conditions
that have made this work possible.
The success of this project also relies critically on 
the dedication of the computing organizations that 
support {\sc BaBar}.
We wish to thank SLAC for its support and the kind hospitality extended to us. 
This work is supported by the
US Department of Energy
and National Science Foundation, the
Natural Sciences and Engineering Research Council (Canada),
Institute of High Energy Physics (China), the
Commissariat \`a l'Energie Atomique and
Institut National de Physique Nucl\'eaire et de Physique des Particules
(France), the
Bundesministerium f\"ur Bildung und Forschung and
Deutsche Forschungsgemeinschaft
(Germany), the
Istituto Nazionale di Fisica Nucleare (Italy),
the Research Council of Norway, the
Ministry of Science and Technology of the Russian Federation, and the
Particle Physics and Astronomy Research Council (United Kingdom). 
Individuals have received support from 
the A. P. Sloan Foundation, the Research Corporation,
and the Alexander von Humboldt Foundation.
The authors wish to thank INFN Sezione di Padova as well as the 
inclusive hadronic $B$ decays analysis working group conveners for their 
support of this analysis. This work is supported by the U.S. Department of Energy 
under grant No. DE-FG05-91ER40622.

%% spires list
%%CITATION = PHRVA,D66,052003;%%
%%CITATION = PHLTA,B482,15;%%
%%CITATION = ZEPYA,C74,19;%%
%%CITATION = PHLTA,B324,249;%%
%%CITATION = PRLTA,89,011802;%%
%%CITATION = PHRVA,D66,010001;%%
%%CITATION = PHRVA,D65,032001;%%
%%CITATION = PRLTA,86,2737;%%
%%CITATION = PHRVA,D42,3885;%%
%%CITATION = PHRVA,D42,3251;%%
%%CITATION = PHRVA,D41,1736;%%
%%CITATION = PHRVA,D21,203;%%
%%CITATION = PRLTA,90,142001;%%
%%CITATION = MPLAE,A18,1783;%%
%%CITATION = PHLTA,B192,245;%%
%%CITATION = PHLTA,B186,247;%%
%%CITATION = PRLTA,74,2626;%%
%%CITATION = PRLTA,74,2632;%%
%%CITATION = ANYAA,578,237;%%
%%CITATION = NUIMA,A342,292;%%
%%CITATION = PRLTA,58,1818;%%
%%CITATION = PHLTA,B139,320;%%
%%CITATION = NUIMA,A316,217;%%
%%CITATION = NUIMA,A420,162;%%
%%CITATION = NUIMA,A502,67;%%
%%CITATION = NUIMA,A494,455;%%
%%CITATION = NUIMA,A479,1;%%
%%CITATION = PRLTA,41,1581;%%
%%CITATION = PRLTA,56,549;%%
%%CITATION = PRLTA,62,1717;%%
%%CITATION = PHLTA,5,165;%%
%%CITATION = PTPSA,37,21;%%
%%CITATION = PHRVA,D67,032002;%%


\begin{references}  % All references should follow standard format
\bibitem{godang02}
CLEO Collaboration, S. B. Athar {\it et al.}, Phys. Rev. D {\bf 66}, 052003 (2002);\\
OPAL Collaboration, G. Abbiendi {\it et al.}, Phys. Lett. B {\bf 482}, 15 (2000);\\
DELPHI Collaboration, P. Abreu {\it et al.}, Z. Phys. C {\bf 74}, 19 (1997);\\
ARGUS Collaboration, H. Albrecht {\it et al.}, Phys. Lett. B {\bf 324}, 249 (1994);\\
{\sc BaBar} Collaboration, B. Aubert {\em et al.}, Phys. Rev. Lett. {\bf 89}, 011802 (2002).
%
\bibitem{pdg2002}
Particle Data Group, K. Hagiwara {\it et al}., Phys. Rev. D {\bf 66}, 010001 (2002).
%
\bibitem{silvia}
{\sc BaBar} Collaboration, B.\ Aubert {\em et al.}, Phys. Rev. D {\bf 65}, 032001 (2002);\\
CLEO Collaboration, J. P. Alexander {\it et al}., Phys. Rev. Lett. {\bf 86}, 2737, (2001).
%
\bibitem{eichten}
N. Byers and E. Eichten, Phys. Rev. D {\bf 42}, 3885 (1990);\\
G. P. Lepage, Phys. Rev. D {\bf 42}, 3251 (1990);\\
D. Atwood and W. J. Marciano, Phys. Rev. D {\bf 41}, 1736 (1990);\\
E. Eichten, K. Gottfried, T. Kinoshita, K. D. Lane,  Phys. Rev. D {\bf 21}, 
203 (1980);\\
R. Kaiser,  A. V. Manohar, and T. Mehen, Phys. Rev. Lett. {\bf 90}, 142001 (2003);\\
M. B. Voloshin, Mod. Phys. Lett. A {\bf 18}, 1783 (2003).
%
\bibitem{argus87}
ARGUS Collaboration, H. Albrecht {\it et al.}, Phys. Lett. B {\bf 192}, 245 (1987).
%
\bibitem{ua187}
UA1 Collaboration, C. Albajar {\it et al.}, Phys. Lett. B {\bf 186}, 247 (1987).
%
\bibitem{cdf95}
CDF Collaboration, F. Abe {\it et al.}, Phys. Rev. Lett. {\bf 74}, 2626 (1995);\\ 
D0 Collaboration, S. Abachi {\it et al.,} Phys. Rev. Lett. {\bf 74}, 2632 (1995). 
%
\bibitem{oddone}
P. Oddone, Annals N.Y. Acad. Sci. {\bf 578}, 237 (1989).
%
\bibitem{burchat94}
{\sc BaBar} Collaboration, P. Burchat, Nucl. Instrum. Meth. A {\bf 342}, 292 (1994);\\ 
E691 Collaboration, J. C. Anjos {\it et al.}, Phys. Rev. Lett. {\bf 58}, 1818 (1987);\\
ACCMOR Collaboration, R. Bailey {\it et al.}, Phys. Lett. B {\bf 139}, 320 (1984). 
%
\bibitem{burchat92}
{\sc BaBar} Collaboration, P. Burchat, J. Hiser, A. Boyarski, and D. Briggs, 
Nucl. Instrum. Meth. A {\bf 316}, 217 (1992). 
%
\bibitem{barlow99}
{\sc BaBar} Collaboration, R. J. Barlow {\it et al.}, Nucl. Instrum. Meth. A {\bf 420}, 162 (1999). 
%
\bibitem{schwiening03}
{\sc BaBar} Collaboration, J. Schwiening {\it et al.}, Nucl. Instrum. Meth. A {\bf 502}, 67 (2003). 
%
\bibitem{anulli02}
{\sc BaBar} Collaboration, F. Anulli {\it et al.}, Nucl. Instrum. Meth. A {\bf 494}, 455 (2002). 
%
\bibitem{nim}
{\sc BaBar} Collaboration, B.\ Aubert {\em et al.}, Nucl. Instrum. Meth. A {\bf 479}, 1 (2002).
%
\bibitem{physbook}
{\sc BaBar} Collaboration, P.\ F.\ Harrison and H.\ R.\ Quinn, ed.,
``The {\sc BaBar} Physics Book,'' {\bf SLAC-R-504} (1998).
%
\bibitem{wolfram}
G. C. Fox and S. Wolfram, Phys. Rev. Lett. {\bf 41}, 1581 (1978).
%
\bibitem{argus86}
ARGUS Collaboration, H. Albrecht {\it et al.}, Phys. Rev. Lett. {\bf 56}, 549 (1986);\\
E691 Collaboration, J. C. Anjos {\it et al.}, Phys. Rev. Lett. {\bf 62}, 1717 (1989).
%
\bibitem{chris}
{\sc BaBar} Collaboration, B.\ Aubert {\em et al.}, Phys. Rev. D {\bf 67}, 032002 (2003).
%
\end{references}
\end{document}